\documentclass[11pt,twoside]{article}
\usepackage{latexsym}
\usepackage[english]{babel}
\usepackage{amssymb,amsfonts,amsmath}
\usepackage{mathrsfs}
\usepackage{fancyhdr}
\usepackage{feynmp}
\usepackage{graphicx}

\newtheorem{theorem}{Theorem}[section]

\newtheorem{lemma}[section]{Lemma}
\newtheorem{definition}[section]{Definition}

\begin{document}

\title{Weak-Coupling Limit. II \\ On the Quantum Fokker-Planck Equation}

\author{David Taj\\ {\small Dept. Physics, Politecnico di Torino, C.so Duca degli
Abruzzi 24, 10129, Torino, Italy} \\ david.taj@gmail.com}

\maketitle

\begin{abstract}
In a recent work we have found a contraction semigroup able to correctly approximate a projected and perturbed one-parameter group of isometries in a generic Banach space, in the limit of weak-coupling. Here we study its generator by specializing to $W^*$-algebras: after defining a Physical Subsystem in terms of a completely positive projecting conditional expectation, we find that it generates a Quantum Dynamical Semigroup. As a consequence of uniqueness and strong generality (well defined dynamics, irrespective of the Physical Subsystem spectral properties or dimensions), its generator deserves to be referred as "the" Quantum Fokker-Planck Equation. We then provide important examples of the limit dynamics, one of which constitutes a new Quantum generalization of the celebrated Fermi Golden Rule.
\end{abstract}

\section*{Introduction}
Recently we have studied some weakly-perturbed one-parameter groups of isometries, projected on Banach subspaces~\cite{tajwbcmp1}. We have found that the projected evolution could be described by a contraction semigroup, under fairly general hypotheses. In particular, no assumption on the subsystem dimensions or spectral properties was made, generalising the results in~\cite{davies1,davies2,spohn}.

For many physical applications however, having a contraction semigroup is certainly not sufficient, as there exists another key fundamental condition that any quantum system must satisfy: that of positivity of the state evolution. Up to date, the need for a correct markovian approximation of a coherent global dynamics, that is able to guarantee the state positivity at all times, can be found in numerous physical contests, such as that of modeling quantum devices~\cite{rossi}, ultrafast spectroscopy in semiconductors~\cite{RMP}, phase transitions~\cite{sachdev}, continuous variable quantum information~\cite{cont_var1,cont_var2} and  quantum open systems~\cite{open,breuerpra,vacchini,breuer} to name a few.

Here we continue the study of the contraction semigroup we have found in our previous work, by specializing the Banach spaces to be generic $W^*$-algebras, in order to be able to address the problem of positivity for the evolution in the most general possible contest (our results can be extended to $C^*$-algebras without essential changes to our presentation).

It turns out that a key definition to get to our result concerns the very same idea of what a Physical Subsystem is: we should define it to be a projecting completely positive normal conditional expectation on a $W^*$-subalgebra. This is also interesting from the mathematical point of view itself, as it is a generalization to the non-commutative case of more the standard concept of conditional probability, and has long been shown to be strictly linked with the basic structures of the involved algebra, such as its modular group of automorphisms~\cite{golodets}.

As a result, we shall prove that the foretold Contraction Semigroup becomes a Quantum Dynamical Semigroup~\cite{lindblad} in case the projected Banach subspace is in fact a Physical Subsystem. This of course is extremely important because it guarantees a positive evolution at all times, together with trace conservation, and thus it becomes immediately applicable to the urgent applicative problems cited above.

After giving a fairly general class of Physical Subsystems, according to our definition, we will discuss and report the explicit limit dynamics of two important examples. The first concerns two or more weakly interacting quantum sectors in a closed setting, contrary to the tensor product structure of open quantum systems. Dissipation here will in fact be possible because information flows irreversibly from the sectors to their polarization space. In the limit of an infinite number of quantum sectors, each sector becomes classically described by a single (positive) occupation probability, and transition rate operators between different sectors boil down to the celebrated Fermi Golden Rule~\cite{FGR}. This motivates the name "Quantum Fermi's Golden Rule" (QFGR) for our example, as the associated dynamics is not a classical Fokker-Planck Equation~\cite{fokker}, but rather a Quantum Dynamical Semigroup~\cite{lindblad}.

The second example is the partial trace over a particle reservoir at thermal equilibrium. This example is new in that we present for the first time the limit dynamics which includes first order contributions, in case no assumption on the subsystem spectral properties are made. This example is easily generalized to the case of some recently proposed "entangling" projections
\cite{breuerpra,vacchini}, through what is reported in our previous results in~\cite{tajpra}.

\section{General Framework}
We report here the essential features of the general framework we'll
be involved with, in order to introduce the basic objects we shall use in the sequel. We suppose that $P_0$ is a linear projection on
a Banach space $\mathcal{B}$,
put $P_1=1-P_0$ and $\mathcal{B}_i=P_i\mathcal{B}$, so that
\begin{equation}
\mathcal{B}=\mathcal{B}_0\oplus\mathcal{B}_1,
\end{equation}
We suppose that $Z$ is the (densely defined) generator of a strongly
continuous one-parameter group of isometries $U_t$ on
$\mathcal{B}$ with
\begin{equation}
U_t P_0=P_0 U_t
\end{equation}
for all $t\in\mathbb{R}$, or equivalently
\begin{equation}
[Z,P_0]=0
\end{equation}
and put $Z_i=P_i Z$. We suppose that $A$ is a bounded perturbation
of $Z$ and put $A_{ij} = P_iAP_j$. We let $U^\lambda_t$ be the one
parameter group generated by $(Z + \lambda A_{00} + \lambda
A_{11})$, and let $V_t^\lambda$ be the one
parameter group generated by $Z+\lambda A$.
Then putting
\begin{equation}
X^\lambda_t=P_0 U^\lambda_t
\end{equation}
and defining the projected evolution as
\begin{equation}
W^\lambda_t=P_0 V^\lambda_t P_0
\end{equation}
and one obtains the all important closed and exact integral
equation
\begin{equation}\label{eq:exact}
W^\lambda_t=X^\lambda_t+\lambda^2 \int_{0}^t ds \int_{0}^s du\;
X^\lambda_{t-s}A_{01}U^\lambda_{s-u} A_{10} W^\lambda_u .
\end{equation}
This is nothing but the integrated form of the well known master
equation constructed by Nakajima, Prigogine, Resibois, and Zwanzig
\cite{naka,zwanzig}.

To recall the basic results we had in~\cite{tajwbcmp1}, we give the following
\begin{definition}\label{def:K_T}
For any real positive $T>0$ put
\begin{equation}
K_T =  {1\over \sqrt{\pi}T} \int_{-\infty}^\infty  dt_1\, e^{-{t_1^2
\over 2 T^2}}\; A_{01}(t_1) \int_{-\infty}^{t_1} dt_2\, e^{-{t_2^2
\over 2 T^2}}\; A_{10}(t_2)
\end{equation}
where we have denoted $A_{ij}(t)=U_{-t} A_{ij} U_{t}$.
\end{definition}
Then we have proven the following
\begin{theorem}\label{th:K_T}
Suppose that $X^\lambda_t$ is a one-parameter group of isometries.
Suppose that there exists some $0<c<\infty$ such that for every
$\overline{\tau}>0$
\begin{equation}\label{eq:no_oscillation_lambda_second}
\int_0^{\lambda^{-2}\overline{\tau}} \|A_{01}U^\lambda_x A_{10}\|
\, dx \leq c
\end{equation}
uniformly on $|\lambda|\leq 1$. Suppose also that for every
$0<\overline{\tau}<\infty$
\begin{equation}\label{hp:lambda_convergence_second}
\lim_{\lambda\rightarrow 0} \int_0^{\lambda^{-2}\overline{\tau}}
\|A_{01}(U^\lambda_x - U_x) A_{10}\| \, dx = 0 \;.
\end{equation}
Let $T\in\mathcal{C}([-1,1],\overline{\mathbb{R}})$ be a real valued positive
continuous function on the interval $[-1,1]$, such that
\begin{equation}
T(\lambda)\sim |\lambda|^{-\xi} \widetilde{T} ,\quad
\lambda\sim 0
\end{equation}
for some real positive reference time $\widetilde{T}>0$ and
scaling $0<\xi<2$. Denote with
\begin{equation}\label{def:semigr_T}
\widetilde{W}^\lambda_t=\exp\{(Z_0+\lambda A_{00}+\lambda^2
K_{T(\lambda)})t\}
\end{equation}
the associated semigroup on $\mathcal{B}_0$.

Then for every $\overline{\tau}>0$
\begin{equation}
\lim_{\lambda\rightarrow 0} \left\{ \sup_{0\leq t\leq
\lambda^{-2}\overline{\tau}}
\|W^\lambda_t-\widetilde{W}^\lambda_t\|\right\}=0.
\end{equation}
\end{theorem}
Our second important result in~\cite{tajwbcmp1} was
\begin{theorem}\label{thr:contraction}
If $\|P_0\|=1$, then $\widetilde{W}^\lambda_t$ is a contraction semigroup on $\mathcal{B}_0$, for all real $\lambda$.
\end{theorem}

\section{The Quantum Fokker-Planck Equation}
In this section we will address the problem of positivity concerning the semigroup $\widetilde{W}^\lambda_t$, that approximates the exact projected evolution (\ref{eq:exact}) in the weak coupling limit. In order to do that, we will restrict our attention to the case $\mathcal{B}=\mathcal{A}$ is a $W^*$-algebra $(\mathcal{A},\mathcal{A}_*)$ with identity,
$\mathcal{A}_*$ being the predual, and $\mathcal{B}_0=\mathcal{X}$ is a $W^*$-subalgebra with identity~\cite{sakai}. However, we observe that all the results of this section could easily be formulated in the more general $C^*$-algebraic contest. We start by introducing some fundamental concepts we shall need in the sequel.\\

Let $\Phi:\mathcal{A}\rightarrow \mathcal{B}$ be a linear map
between two $C^*$-algebras $\mathcal{A}$ and $\mathcal{B}$. Let
$M_n(\mathcal{A})$ denote the $n\times n$-matrix algebra over
$\mathcal{A}$ (same for $\mathcal{B}$). Then $\Phi$ induces a map
$\Phi_n:M_n(\mathcal{A})\rightarrow M_n(\mathcal{B})$ defined by
\begin{equation}
\Phi_n(X\otimes E_{ij})=\Phi(X)\otimes E_{ij},
\end{equation}
where $E_{ij}, i,j=1\ldots n$ are the matrix units spanning
$M_n(\mathbb{C})$.
\begin{definition}
$\Phi:\mathcal{A}\rightarrow \mathcal{B}$ is said to be positive
if for every $X\in\mathcal{A}$, $\Phi(X^\dagger X)=Y^\dagger Y$
for some $Y\in\mathcal{B}$. Equivalently, $\Phi$ is positive if
$\Phi(A)$ is positive whenever $A$ is positive.
$\Phi:\mathcal{A}\rightarrow \mathcal{B}$ is said to be completely
positive iff $\Phi_n$ is positive for all $n$.
\end{definition}
For sake of completeness we report from \cite{lindblad} the following
\begin{definition}
Let $\mathcal{X}$ be a $W^*$-algebra with identity. A Quantum Dynamical Semigroup (QDS) is a one-parameter family of maps $\Phi_t$ of $\mathcal{X}$ into itself satisfying
\begin{enumerate}
\item[i)] $\Phi_t$ is completely positive;

\item[ii)] $\Phi_t(1)=1$;

\item[iii)] $\Phi_s \Phi_t=\Phi_{s+t}$;

\item[iv)] $\Phi_t(X)\rightarrow X$ ultraweakly, $t\rightarrow 0$, $\forall X\in\mathcal{X}$;

\item[v)] $\Phi_t$ is normal (ultraweakly continuous).
\end{enumerate}
\end{definition}
Suppose that $L:\mathcal{X}\rightarrow\mathcal{X}$ is of the form
\begin{equation}\label{def:lindblad_form}
L(X)=i[H,X]-{1\over 2}\{A,X\} + \Psi(X),
\end{equation}
where $H$ is a (unbounded) self-adjoint operator on $\mathcal{X}$, $A\in\mathcal{X}$ is self-adjoint, $\Psi:\mathcal{X}\rightarrow\mathcal{X}$ is completely positive and $\Psi(1)=A$. Then it is very well known \cite{lindblad} that $L$ generates a Quantum Dynamical Semigroup through $\Phi_t=\exp\{Lt\}$.
\\

Now let $\mathcal{X}$ be a $W^*$-subalgebra of $\mathcal{A}$: we have the following diagram
\begin{equation}
\begin{array}{ccc}
\mathcal{X} & \hookrightarrow & \mathcal{A} \\
\downarrow  &                 & \downarrow \\
\mathcal{X}_*& \leftarrow & \mathcal{A}_*
\end{array}
\end{equation}
where $\downarrow$ represent the dualities $\langle\cdot,\cdot\rangle$, horizontal arrows are the inclusion and quotient map, $\mathcal{X}_*= \mathcal{A}_* / \mathcal{X}_0$ and $\mathcal{X}_0$ is the polar of $\mathcal{X}$ in $\mathcal{A}_*$, defined as
\begin{equation}
\mathcal{X}_0=\{\rho\in\mathcal{A}_* \;|\; \langle X,\rho \rangle =0 \quad\forall X\in\mathcal{X} \}.
\end{equation}
We report here a slight modification of what is found in~\cite{davies3} in that we require complete positivity:
\begin{definition}\label{def:cond_exp}
Let $\mathcal{X}$ be a $W^*$-subalgebra of a $W^*$-algebra
$\mathcal{A}$. A completely positive normal conditional expectation of $\mathcal{A}$ onto
$\mathcal{X}$ is a linear map $\Phi:\mathcal{A}\rightarrow
\mathcal{X}$ such that
\begin{itemize}
\item $\Phi(X^\dagger)=\Phi(X)^\dagger$;

\item $\Phi(X)=X$ if and only if $X\in\mathcal{X}$;

\item $\Phi$ is completely positive;

\item if $X_1,X_2\in\mathcal{X}$ and $Y\in\mathcal{A}$, then
$\Phi(X_1YX_2)=X_1\Phi(Y)X_2$

\item $\Phi(X_n)\uparrow\Phi(X)$ whenever $X_n\uparrow X$ ultraweakly
\end{itemize}
(the term "normal" or "ultraweakly continuous" refers to the this last requirement).

We further define a completely positive projecting normal conditional expectation (CPPNCE) to be a completely positive normal conditional expectation, which is also a projection.
\end{definition}
We shall from now on denote the projected $P_0(X)$ with the
expectation symbol $P_0(X)=\langle X \rangle$, or $P_0=\langle \cdot \rangle$, because of our following
\begin{definition}\label{def:subsystem}
A Physical Subsystem $\mathfrak{X}$ is a triple $\mathfrak{X}=(\mathcal{X},\mathcal{A},\langle\cdot\rangle)$ where $\mathcal{A}$ is a $W^*$-algebra, $\mathcal{X}\hookrightarrow\mathcal{A}$ is a $W^*$-subalgebra with identity, and $\langle\cdot\rangle:\mathcal{A}\rightarrow \mathcal{X}$ is a CPPNCE.
\end{definition}

Then under suitable natural (and fairly general) hypotheses, we shall show in this section that the semigroup $\widetilde{W}^\lambda_t$, defined on a Physical Subsystem $\mathfrak{X}=(\mathcal{X},\mathcal{A},\langle\cdot\rangle)$, is a Quantum Dynamical Semigroup.

To state the main result of this section, we shall need the following
\begin{definition}
For any real $\omega$, the $\omega$-translated dynamically coarse-grained perturbation $\mathcal{L}_{\lambda\omega}$ associated to $H'\in\mathcal{A}$ is given by
\begin{equation}
\mathcal{L}_{\lambda\omega}= \sqrt{1\over \sqrt{\pi}T(\lambda)}
\int_{-\infty}^\infty dt\, e^{i\omega t} \,e^{-{t^2 \over 2
T(\lambda)^2}}\; H'(t).
\end{equation}
We shall refer to $\mathcal{L}_\lambda=\mathcal{L}_{\lambda 0}$
simply as dynamically coarse-grained perturbation.
\end{definition}
\begin{theorem}\label{th:QDS}
Let $\mathfrak{X}=(\mathcal{X},\mathcal{A},\langle\cdot\rangle)$ be a Physical Subsystem.

Let $U_t=\exp\{Z t\}$ be a one-parameter group of automorphisms on $\mathcal{A}$ generated by an (unbounded) self-adjoint operator $H_0$ (formally) through
\begin{equation}
Z(X)=i[H_0,X]
\end{equation}
and assume that $\langle H_0 \rangle$ is an (unbounded) self-adjoint operator on $\mathcal{X}$.

Suppose $A$ is a bounded self-adjoint derivation on $\mathcal{A}$.

Then
\begin{enumerate}
\item[i)] $X^\lambda_t=e^{(Z_0+\lambda A_{00})t}$ is a one-parameter group of automorphisms on $\mathcal{X}$.

\item[ii)] $\widetilde{W}_t^\lambda=\exp\{(Z_0+\lambda A_{00}+\lambda^2
K_{T(\lambda)})t\}$ is a Quantum Dynamical Semigroup.

\item[iii)] Moreover, its generator has the form
\begin{eqnarray}
\partial_t X &=& i[\langle H_\lambda \rangle,X] + i \left[\int {d\omega \over 2\pi \omega}
\langle (\mathcal{L}_{\lambda \omega}- \langle\mathcal{L}_{\lambda \omega}\rangle)^\dagger
(\mathcal{L}_{\lambda \omega}- \langle\mathcal{L}_{\lambda \omega}\rangle) \rangle,X\right] \nonumber \\
&& - {1\over 2} \{\langle (\mathcal{L}_\lambda  - \langle
\mathcal{L}_\lambda \rangle)^2 \rangle, X \} + \langle
(\mathcal{L}_\lambda-\langle \mathcal{L}_\lambda \rangle) X
(\mathcal{L}_\lambda-\langle \mathcal{L}_\lambda \rangle)\rangle.
\end{eqnarray}
\end{enumerate}
where the ($\omega$-translated) dynamically coarse-grained perturbations $\mathcal{L}_{\lambda \omega}$ are associated to a uniquely defined, up to addition of the identity\footnote{Note that substituting $H'\mapsto H'+\alpha 1$ leaves the generator unaffected}, self-adjoint element $H'\in\mathcal{A}$, and we have put
\begin{equation}
H_\lambda=H_0+\lambda H'.
\end{equation}
\end{theorem}

{\bf Proof}.
First we know \cite{sakai} that every bounded derivation of a $W^*$-algebra is inner. Hence there is a
$Y\in \mathcal{A}$ such that $A(X)=[Y,X]$. But from
$A(X^\dagger)=A(X)^\dagger$ it follows that $Y=i H'$ for a
self-adjoint element $H'\in \mathcal{A}$, so that
\begin{equation}
A(X)=i[H',X].
\end{equation}
To prove $i)$ note that $\langle H' \rangle$ is self-adjoint, as $H'$ is self-adjoint and $\langle\cdot\rangle$ is an adjoint map. Then $\langle H_\lambda \rangle = \langle H_0 \rangle +\lambda \langle H' \rangle$ is a (unbounded) self-adjoint operator on $\mathcal{X}$ and
\begin{equation}
X^\lambda_t(X)=e^{i[\langle H_\lambda \rangle,\cdot]t}(X)=e^{i \langle H_\lambda \rangle t} X e^{-i \langle H_\lambda \rangle t}
\end{equation}
is a one-parameter group of automorphisms on $\mathcal{X}$.

The validity of $ii)$ follows from $iii)$, by just noting that equation $iii)$ is in the Lindblad form (\ref{def:lindblad_form}). In fact, both $X\mapsto \langle X\rangle$ and $X\mapsto (\mathcal{L}_\lambda-\langle \mathcal{L}_\lambda \rangle) X
(\mathcal{L}_\lambda-\langle \mathcal{L}_\lambda \rangle)$ are completely positive maps (the latter is competely positive since it has the Kraus form \cite{kraus}), and so is their composition \cite{lindblad}. The remaining requirements $A\in\mathcal{X}$ self-adjoint and $\Psi(1)=A$ following (\ref{def:lindblad_form}) can easily be checked.

To show $iii)$, we start putting $K_{T(\lambda)}$ in a more convenient form: we name
\begin{equation}
\Phi^\lambda(t)=\sqrt{1\over \sqrt{\pi}T(\lambda)}\; e^{-{t^2\over 2T(\lambda)^2}}\; U_{-t}A U_t
\end{equation}
and denote as usual $\Phi^\lambda_{ij}(t)=P_i\Phi^\lambda(t) P_j$. Then from
\begin{eqnarray}
K_{T(\lambda)} &=& \int_{-\infty}^{+\infty} \!\!\! dt_1 \int_{-\infty}^{+\infty}\!\!\! dt_2\; \Phi^\lambda_{01}(t_1) \Phi^\lambda_{10}(t_2) -\int_{-\infty}^{+\infty} \!\!\! dt_1\int_{-\infty}^{t_1} \!\!\! dt_2\;  \Phi^\lambda_{01}(t_2) \Phi^\lambda_{10}(t_1) \nonumber\\
K_{T(\lambda)} &=&  \int_{-\infty}^{+\infty} \!\!\! dt_1 \int_{-\infty}^{t_1}\!\!\! dt_2 \; \Phi^\lambda_{01}(t_1) \Phi^\lambda_{10}(t_2)
\end{eqnarray}
we sum term by term to obtain
\begin{equation}
K_{T(\lambda)} = {1\over 2} \int_{-\infty}^{+\infty} \!\!\! dt_1 \; \Phi^\lambda_{01}(t_1) \int_{-\infty}^{+\infty} \!\!\! dt_2 \; \Phi^\lambda_{10}(t_2) + {1\over 2} \int_{-\infty}^{+\infty} \!\!\! dt_1 \int_{-\infty}^{t_1} \!\!\! dt_2\;  [\Phi^\lambda_{01}(t_1), \Phi^\lambda_{10}(t_2)].
\end{equation}
We now introduce
\begin{eqnarray}
\widetilde{K}_{T(\lambda)} &=& {1\over 2} \int_{-\infty}^{+\infty} \!\!\! dt_1 \; (\Phi^\lambda-\Phi^\lambda_{00})(t_1) \int_{-\infty}^{+\infty} \!\!\! dt_2 \; (\Phi^\lambda-\Phi^\lambda_{00})(t_1) \\
&+& {1\over 2}\int_{-\infty}^{+\infty} \!\!\! dt_1 \int_{-\infty}^{+\infty} \!\!\! dt_2\;  (\Phi^\lambda-\Phi^\lambda_{00})(t_1) (\Phi^\lambda-\Phi^\lambda_{00})(t_2) \; \rm{sign}(t_1-t_2),\nonumber
\end{eqnarray}
where $(\Phi^\lambda-\Phi^\lambda_{00})(t)=\Phi^\lambda(t)-\Phi^\lambda_{00}(t)$ and
\begin{equation}
\rm{sign}(t)=\left\{
\begin{array}{cc}
1, & t> 0 \\
0, & t=0 \\
-1, & t<0
\end{array}
\right.
\end{equation}
is the sign function. Then it follows easily that
\begin{equation}
K_{T(\lambda)}=P_0\widetilde{K}_{T(\lambda)} P_0.
\end{equation}
Now we write the Fourier representation
\begin{equation}
\rm{sign}(t)=i\int {d\omega\over\pi\omega}\; e^{-i\omega t},
\end{equation}
and introduce notation
\begin{equation}
\mathbb{L}_{\lambda,\omega}=\int_{-\infty}^{+\infty} \!\!\! dt_1 \; e^{i\omega t}(\Phi^\lambda-\Phi^\lambda_{00})(t)
\end{equation}
and $\mathbb{L}_\lambda=\mathbb{L}_{\lambda 0}$, so that
we obtain
\begin{equation}\label{eq:K_T_positive_general}
K_{T(\lambda)} = {1\over 2}\, P_0 \mathbb{L}_\lambda^2 P_0+ i\int {d\omega\over2\pi\omega}\,P_0 \mathbb{L}_{\lambda,-\omega}\mathbb{L}_{\lambda,\omega} P_0
\end{equation}
In order to compute $K_{T(\lambda)}$, let's also note that for every
$X,Y\in\mathcal{A}$, and real $t$,
\begin{equation}
U_t(XY)=U_t(X)U_t(Y)
\end{equation}
since $U_t$ is an automorphism. Then it follows easily that
\begin{equation}
A_{ij}(t)(X)= U_{-t}A_{ij}U_t(X)= iP_i( [H'(t),P_j(X)])
\end{equation}
where in the last line we have defined the interaction picture
hamiltonian $H'(t)=U_{-t}(H')$.

We will now consider each factor
in (\ref{eq:K_T_positive_general})
separately. Using the definition of the
$\omega$-translated dynamically coarse-grained perturbation operators
$\mathcal{L}_{\lambda\omega}$, we recognise that
\begin{equation}
\int_{-\infty}^{+\infty} \!\!\! dt \; (\Phi^\lambda-\Phi^\lambda_{00})(t)\;(X) = i[\mathcal{L}_\lambda,X] -i\langle[\mathcal{L}_\lambda, \langle X\rangle]\rangle, \qquad X\in\mathcal{A},
\end{equation}
so we take some $X=\langle X\rangle\in\mathcal{X}$ and
compute
\begin{eqnarray}
P_0\mathbb{L}_\lambda^2 P_0 (X) &=& \langle
[\mathcal{L}_\lambda, [\mathcal{L}_\lambda, X]] \rangle - \langle
[\mathcal{L}_\lambda, \langle [\mathcal{L}_\lambda, X]] \rangle
\rangle \nonumber\\
&=& \langle [\mathcal{L}_\lambda, [\mathcal{L}_\lambda, X]]
\rangle -  [\langle \mathcal{L}_\lambda \rangle, [ \langle
\mathcal{L}_\lambda \rangle, X]]
\end{eqnarray}
The second line follows because $\langle\cdot\rangle$ is a conditional
expectation, and so for example $\langle [\mathcal{L}_\lambda,
X]\rangle=[\langle \mathcal{L}_\lambda \rangle, X]$ (recall that
$X\in\mathcal{X}$). Expanding the commutators, this can be
written as
\begin{eqnarray}
P_0\mathbb{L}_\lambda^2 P_0 (X) &=& - \{\langle(\mathcal{L}_\lambda-\langle \mathcal{L}_\lambda \rangle)^2\rangle, X \} + 2 \langle
(\mathcal{L}_\lambda-\langle \mathcal{L}_\lambda \rangle) X
(\mathcal{L}_\lambda-\langle \mathcal{L}_\lambda \rangle)\rangle.
\end{eqnarray}
The second factor in
(\ref{eq:K_T_positive_general}) applied to some $X\in\mathcal{X}$, and projected,
can be treated with similar calculations to find
\begin{equation}
P_0\mathbb{L}_{\lambda,-\omega}\mathbb{L}_{\lambda,\omega}P_0 X =
\left[\langle (\mathcal{L}_{\lambda \omega}- \langle\mathcal{L}_{\lambda \omega}\rangle)^\dagger
(\mathcal{L}_{\lambda \omega}- \langle\mathcal{L}_{\lambda \omega}\rangle) \rangle, X\right]
\end{equation}
Putting the results together, we have computed
\begin{eqnarray}
Z_0 X &=& i[\langle H_0 \rangle,X] \nonumber \\
A_{00} X &=&  i[\langle H' \rangle,X]\nonumber \\
K_{T(\lambda)} X &=& i \left[ \int {d\omega \over 2\pi \omega}
\langle (\mathcal{L}_{\lambda \omega}- \langle\mathcal{L}_{\lambda \omega}\rangle)^\dagger
(\mathcal{L}_{\lambda \omega}- \langle\mathcal{L}_{\lambda \omega}\rangle) \rangle, X\right] \nonumber \\
&& - {1\over 2} \{\langle(\mathcal{L}_\lambda-\langle \mathcal{L}_\lambda \rangle)^2\rangle, X \}+ \langle (\mathcal{L}_\lambda-\langle \mathcal{L}_\lambda
\rangle) X (\mathcal{L}_\lambda-\langle \mathcal{L}_\lambda
\rangle)\rangle,
\end{eqnarray}
showing $iii)$ and completing the proof.
$\quad\Box$ \\

\subsection{Comments to the theorem}
\begin{itemize}
\item Note that, put in this form, $K_{T(\lambda)}$ furnishes a
dynamical measure of how much the CPPNCE $\langle\cdot\rangle$ differs from
being an algebra homomorphism (as $K_{T(\lambda)}=0$ in that
case), thus giving dynamical information on the nature of the
Physical Subsystem $\mathfrak{X}$.

\item $\|P_0\|=1$, in agreement with our previous requirements for the validity of Theorem \ref{thr:contraction}. To show this, note that $P_0$ is a completely positive map on a $C^*$-algebra and so, according to ~\cite{stormer},
\begin{equation}
P_0(X^\dagger)P_0(X)\leq P_0(X^\dagger X).
\end{equation}
Passing to the norms in case $X=U$ is unitary we obtain (recall $P_0$ is an adjoint map) $\|P_0(U)\|^2\leq \|P_0(1)\|=\|1\|=1$.
But then from Corollary 1 in~\cite{russo} we know that $\|P_0\|=\sup_U \|P_0(U)\|$, where the $\sup$ is taken among all unitary operators: since $1_\mathcal{A}$ is unitary, it follows that $\|P_0\|=\|P_0(1)\|=1$.
\end{itemize}
Through the next Lemma, it will turn out that the hypotheses of last theorem are of quite
general nature. Moreover, its results allow the explicit construction of a fairly large number of Physical Subsystems.
\begin{lemma}\label{lemma:positivity}
Let $P_0$ be a projection on $\mathcal{A}$ of the form
\begin{equation}\label{def:cpprojection}
P_0(X)=\sum_{\alpha\in\Lambda} V_\alpha^\dagger X V_\alpha
\end{equation}
for some $\{V_\alpha\}_{\alpha\in\Lambda}\subset\mathcal{A}$ and
some (possibly uncountable) indexing set $\Lambda$. Suppose that
$P_0$ projects onto a $W^*$-subalgebra $\mathcal{X}\hookrightarrow\mathcal{A}$ and that $1_\mathcal{A}\in\mathcal{X}$. Denote
with
\begin{equation}
\mathcal{C}=\{X\in\mathcal{A} \;|\;
[V_\alpha,X]=[V_\alpha^\dagger,X]=0,\;\alpha\in\Lambda\}
\end{equation}
the set of all the elements in $\mathcal{A}$ that commute with
each of the $V_\alpha$ and $V_\alpha^\dagger$.

Then $\mathcal{X}=\mathcal{C}$, and $P_0:\mathcal{A}\rightarrow\mathcal{X}$ is a CPPNCE.
\end{lemma}

{\bf Proof}. It's straightforward to show that $\mathcal{C}\subset
\mathcal{X}$. In fact, let $X\in\mathcal{C}$ and evaluate
\begin{equation}
P_0(X)=\sum_\alpha V_\alpha^\dagger V_\alpha X=X \;,
\end{equation}
(the last equality follows from $P_0(1)=1$, as $1_\mathcal{A}\in\mathcal{X}$ and $P_0$ is a projection), which shows that
$X\in\mathcal{X}=\Im(P_0)$.

Then we show that $Y\in\Im(P_0)=\mathcal{X}\Rightarrow
Y\in\mathcal{C}$. To this end, take any $X\in\mathcal{X}$, so that
$XY\in\mathcal{X}$ because $\mathcal{X}$ is an algebra. Since the
restriction $P_0|_{\mathcal{X}}$ is the identity, we have
\begin{equation}\label{eq:imp}
\sum_\alpha V_\alpha^\dagger XY V_\alpha = XY \;.
\end{equation}
Writing this sum as
\begin{equation}
\sum_\alpha V_\alpha^\dagger XY V_\alpha = \sum_\alpha
V_\alpha^\dagger X V_\alpha Y+\sum_\alpha V_\alpha^\dagger X[Y,
V_\alpha]
\end{equation}
and noting that the first term in the right hand side is nothing
but $XY$, as $X\in\mathcal{X}$ by hypothesis, we conclude that
equality (\ref{eq:imp}) amounts to
\begin{equation}
\sum_\alpha V_\alpha^\dagger X[Y, V_\alpha] = 0\;.
\end{equation}
But this is true for every $X\in\Im (P_0)=\mathcal{X}$, so
\begin{equation}
[Y, V_\alpha] = 0 \quad \forall \alpha \;.
\end{equation}
We can prove $[Y,V_\alpha^\dagger]$ in a perfectly analogous
manner, so this, together with the ultraweak continuity of
(\ref{def:cpprojection}) (see \cite{sakai}), shows that $\mathcal{X}=\mathcal{C}$.

It is then easy to realize that $P_0$ is a projecting normal conditional expectation, by just checking the requirements in Definition \ref{def:cond_exp} and noting that $P_0$ has the Kraus form \cite{kraus} (which guarantees complete positivity).
$\quad\Box$ \\

The example of CPPNCE reported in the last lemma is fairly general, and for example it includes as a special case, as we shall see, the partial tracing over a bath.

Then, to all extent, the last theorem furnishes enough evidence to
consider
\begin{equation}\label{eq:QFP}
\partial_t X = (Z_0+\lambda A_{00}+\lambda^2
K_{T(\lambda)})X
\end{equation}
as "the" Quantum Fokker-Planck Equation. In fact, the equation is
well defined for every nonzero value of the coupling constant
$\lambda$, no matter which are the subsystem's nature (that of $\mathcal{X}$ and $\langle\cdot\rangle$),
spectral properties (that of $Z_0$), or dimensions
(dimension of the subalgebra $\mathcal{X}$ as a vector space). Moreover, it always
gives rise to a (completely positive and trace preserving) Quantum
Dynamical Semigroup $\widetilde{W}^\lambda_t$, which is compatible with the exact evolution $W^\lambda_t$
in the weak-coupling limit (if the convergence hypotheses of Theorem \ref{th:K_T} are satisfied). As such, it clearly generalizes the
preexisting results in \cite{davies1,davies2}.

\section{Quantum Fermi Golden Rule}
The conditions for the validity of the previous theorem are
satisfied for a fairly general class of projections $P_0$. As a
first important example, suppose $\{V_\alpha\}_\alpha$ is a complete set of
mutually orthogonal projections on the Hilbert space $\mathcal{H}$, so that
\begin{equation}
V_\alpha^\dagger=V_\alpha,\quad V_\alpha V_\beta=\delta_{\alpha\beta}V_\beta \quad \rm{and} \quad \sum_\alpha V_\alpha V_\alpha=1.
\end{equation}
Take $(\mathcal{A},\mathcal{A}_*)=(B(\mathcal{H}),\mathcal{T}(\mathcal{H}))$,
the $W^*$-algebra of all bounded operators on $\mathcal{H}$,
together with the ultraweak topology induced by the predual
$\mathcal{T}(\mathcal{H})$, the space of
trace-class operators on $\mathcal{H}$. It is clear that
\begin{equation}
\mathcal{X}=\{X\in B(\mathcal{H}) \;|\; \forall \alpha\; [X,V_\alpha]=0 \}= \left\{X\in B(\mathcal{H}) \;|\; X=\sum_\alpha V_\alpha X V_\alpha \right\}
\end{equation}
is the $W^*$-subalgebra of box-diagonal elements in $B(\mathcal{H})$. The predual Banach space $\mathcal{X}_*$
is readily identified with
\begin{equation}
\mathcal{X}_*=\{\rho\in\mathcal{T}(\mathcal{H}) \; |
\rho=\sum_\alpha V_\alpha \rho V_\alpha\}.
\end{equation}
Define the completely positive projection $P_0(X)=\sum_\alpha V_\alpha X V_\alpha$ on $B(\mathcal{H})$ and note that, because of Lemma \ref{lemma:positivity}, it is a (normal) conditional expectation, so that $\mathfrak{X}=(\mathcal{X},B(\mathcal{H}),P_0)$ is a Physical Subsystem.

Suppose a hamiltonian $H_\lambda=H_0+\lambda H'$ is given on $\mathcal{H}$, with
$[H_0,V_\alpha]=0$ for all $\alpha$: then $[Z,P_0]=0$.
After naming $\rho_\alpha=V_\alpha \rho V_\alpha$ for $\rho\in\mathcal{X}_*$ and $X_\alpha=V_\alpha X V_\alpha$ for $X\in\mathcal{X}$, our Quantum Fokker-Planck equation (\ref{eq:QFP}) becomes, in Schroedinger picture, a coupled set of equations of the form
\begin{eqnarray}\label{eq:qfgr_1order}
\partial \rho_\alpha &=& -i[H_\alpha+\lambda H'_{\alpha} +\lambda^2 H''_{\lambda,\alpha},\rho_\alpha]\nonumber\\
&-&{\lambda^2\over 2}\sum_{\beta\neq\alpha}
\{ D_{\lambda\alpha\beta}^\dagger D_{\lambda\alpha\beta},\rho_\alpha\}
+\lambda^2 \sum_{\beta\neq\alpha} D_{\lambda\alpha\beta} \rho_\beta
D_{\lambda\alpha\beta}^\dagger
\end{eqnarray}
with subsystem hamiltonian $H_\alpha=V_\alpha H_0
V_\alpha=V_\alpha H_0$, first order contribution
$H'_{\alpha}=V_\alpha H' V_\alpha$, second order energy
renormalization
\begin{equation}
H''_{\lambda,\alpha} = \sum_{\beta\neq\alpha}\; \int {d\omega\over 2\pi\omega} \;V_\alpha \mathcal{L}_{\lambda\omega}^\dagger V_\beta \mathcal{L}_{\lambda\omega} V_\alpha
\end{equation}
and scattering operators
\begin{equation}
D_{\lambda\alpha\beta}=V_\beta \mathcal{L}_\lambda V_\alpha.
\end{equation}
The operators $D_{\lambda\alpha\beta}$ can be interpreted as "quantum
transition amplitudes" among the "quantum populations"
$\{\rho_\alpha\}$, that in fact couple the different populations
and guarantee positivity of each, as one
can easily see.

The generator for $\rho\in\mathcal{X}_*$ in (\ref{eq:qfgr_1order}) thus constitutes a coupled linear system for density matrices
$\rho_\alpha$. It is certainly of special interest, as the "transition rates" $D_{\lambda\alpha\beta}^\dagger D_{\lambda\alpha\beta}$ between the density matrices $\rho_\alpha$ and $\rho_\beta$, with hamiltonians $H_\alpha$ and $H_\beta$, furnish a quantum analog of the classical Fermi Golden Rule transition rates \cite{alicki}
\begin{equation}
\mathcal{P}_{\alpha\beta}=2\pi \; H'^\alpha_\beta\, H'^\beta_\alpha\, d\alpha\;\; \delta(\epsilon_\alpha-\epsilon_\beta)
\end{equation}
between $H_0$ eigenstates $|\alpha\rangle$ and $|\beta\rangle$ with energies $\epsilon_\alpha$ and $\epsilon_\beta$\footnote{One normally computes the transition rates for an initial state $|\alpha\rangle$ belonging to the discrete part of the spectrum of $H_0$, while $|\beta\rangle$ belongs to its continuum part: to account also for transitions to happen within the continuous part of $H_0$, one has to multiply the standard transition rate by the spectral measure $d\alpha$ in $H_0=\int d\alpha |\alpha\rangle \epsilon_\alpha \langle\alpha|$, to conserve dimensions and avoid singularities.}. Indeed, in the singular, diagonal case $V_\alpha=|\alpha\rangle\langle\alpha| d\alpha$ (same for $\beta$), one could easily see that
\begin{equation}
\lim_{\lambda\rightarrow 0} D_{\lambda\alpha\beta}^\dagger D_{\lambda\alpha\beta} = \mathcal{P}_{\alpha\beta} \; d\beta \;\; V_\alpha
\end{equation}
recovers the Fermi Golden Rule, with the associated classical Fokker-Plank Equation
\begin{equation}
\partial_t f_\alpha = \lambda^2 \int d\beta\; \left( \mathcal{P}_{\beta\alpha} f_\beta  - \mathcal{P}_{\alpha\beta} f_\alpha \right) .
\end{equation}

Note that this example is peculiar to the case $\mathcal{B}_0$ is not
finite dimensional and $Z_0$ has (also) continuous spectrum. In fact,
only in these conditions one can hope to show that
\begin{equation}
\int_0^\infty \|A_{01}U_t A_{10}\| \, dt < \infty,
\end{equation}
which is a necessary requirement for all the theorems on the weak
limit we have proven. Physically, one could say that only when the
spectrum of the free hamiltonian is continuous, the "off-diagonal polarizations"
$V_\alpha\rho V_\beta$ ($\alpha \neq \beta$) contain enough (in fact an infinite
number of) degrees of freedom, to allow an exponential decay solution
instead of Bloch oscillations. As it is, equation (\ref{eq:qfgr_1order}) could
thus be addressed as a "Quantum Fermi's Golden Rule" (QFGR), as the
"quantum populations" $\rho_\alpha$ are (positive
definite) density matrices rather then (positive) real numbers representing the eigenstates $|\alpha\rangle$ populations.

This deserves a comment: there are many proposed quantum
generalization of the well known Fermi's Golden Rule
\cite{open,FGR}, which are robust and physically and
mathematically meaningful. All these generalizations consider a
bipartite system: nevertheless, the original idea by Fermi,
stated in modern terms, was rather to take a global system
and project on the space of density matrices, that are diagonal in
the basis of the unperturbed hamiltonian. His motivations referred
to the fact that the system eigenvalues are "robust" against
dissipation, and thus constitute the relevant degrees of freedom.
Just along these lines, no environment (system "B") is present in our
model of the QFGR. The possibility to obtain a QDS comes exactly by the fact that the
unperturbed hamiltonian has continuous spectrum, thus conferring
our QFGR version an autonomous relevance.

We believe that our QFGR equation could be applied in numerous topical and important cases in the next future, to study dissipation, decoherence and quantum noise among weakly interacting continuous quantum subsectors of a given system.

\section{Partial Tracing over a Heat Bath}
What we say here for a system "A" coupled to a bath "B" holds for
the general case of Quantum Open Systems, that is, one could easily implement more reservoirs at different temperatures (for an interesting and important generalization of this case see \cite{tajpra} and references therein).

The projection $P_0$ we are going to define is informally referred to as "tracing away the bath
degrees of freedom". Let
$\mathcal{H}=\mathcal{H}_A\otimes\mathcal{H}_B$ be the tensor
product of two Hilbert spaces and denote with
$\mathcal{T}(\mathcal{K})$ the space of trace-class operators on
$\mathcal{K}$. Now
$\widetilde{P}_0:\mathcal{T}(\mathcal{H})\rightarrow\mathcal{T}(\mathcal{H}_A)$
is uniquely determined by
\begin{equation}
\rm{Tr}(\widetilde{P}_0(\rho)X)=\rm{Tr}(\rho(X\otimes 1_B))
\end{equation}
for arbitrary $\rho\in \mathcal{T}(\mathcal{H})$ and bounded
operator $X$ on $\mathcal{H}_A$. If a (normal) state $\omega\geq
0$, $\rm{Tr}(\omega)=1$, is given on $\mathcal{H}_B$, then
$\overline{P}_0\rho=\widetilde{P}_0(\rho)\otimes \omega$ is a
projection in $\mathcal{T}(\mathcal{H})$ with image isomorphic to
$\mathcal{T}(\mathcal{H}_A)$, called the partial trace.

Let $\{\phi_\alpha\}$ be a basis for $\mathcal{H}_B$ and define
the operators
\begin{equation}
V_{\alpha\beta}=1_A\otimes \sqrt{\omega}|\phi_\beta\rangle\langle
\phi_\alpha|
\end{equation}
on $\mathcal{H}$. Then
\begin{equation}\label{eq:trace_cp}
\overline{P}_0\rho=\sum_{\alpha\beta} V_{\alpha\beta} \rho
V_{\alpha\beta}^\dagger.
\end{equation}
Now set $\mathcal{A}=B(\mathcal{H})$, and define $P_0$ on
$\mathcal{A}$ through
\begin{equation}
P_0X=\sum_{\alpha\beta} V_{\alpha\beta}^\dagger X V_{\alpha\beta}
\;.
\end{equation}
It's easy to check that $P_0$ is a
completely positive linear projection, and also an ultraweakly continuous
conditional expectation, on the $W^*$-algebra
$\mathcal{A}=B(\mathcal{H})$ with predual
$\mathcal{A}_*=\mathcal{T}(\mathcal{H})$. Then the $W^*$-subalgebra $\mathcal{X}$ of operators on
$\mathcal{H}$ that commute with each $V_{\alpha\beta}^{(\dagger)}$
(see Lemma \ref{lemma:positivity}) is precisely
\begin{equation}
\mathcal{X}=\{\widetilde{X}\in B(\mathcal{H}) \;|\;
\widetilde{X}=X\otimes 1_B  \;\;\rm{ for\; some }\;\;  X\in
B(\mathcal{H}_A)\}\sim B(\mathcal{H}_A).
\end{equation}
The polar $\mathcal{X}_0$ of $\mathcal{X}$ in
$\mathcal{A}_*=\mathcal{T}(\mathcal{H})$ is thus
\begin{equation}
\mathcal{X}_0=\{\rho\in\mathcal{A}_* \;|\; \rm{Tr} (\rho\; X\otimes
1)=0\;\; \forall\, X\in B(\mathcal{H}_A)\},
\end{equation}
so that from
\begin{equation}
\left.
\begin{array}{c}
\sigma \in \mathcal{T}(\mathcal{H}_B) \\
\rm{Tr}(\sigma) = 1
\end{array}
\right\}
\;\;\Longrightarrow\;\;
\sum_\alpha \rho_\alpha\otimes \sigma_\alpha - \left(\sum_\alpha
\rm{Tr}(\sigma_\alpha) \rho_\alpha\right) \otimes \sigma \in
\mathcal{X}_0
\end{equation}
we see that $\mathcal{X}_*=\mathcal{A}_* / \mathcal{X}_0$ can easily be
identified with $\mathcal{X}_*=\mathcal{T}(\mathcal{H}_A)$.

In particular, we see that the projection $\overline{P}_0$ could
be used to pass to the quotient through $\rho_1\sim\rho_2$ in
$\mathcal{T}(\mathcal{H})$ if and only if
$\overline{P}_0(\rho_1)=\overline{P}_0(\rho_2)$ (alternatively,
one could put $\widetilde{P}_0(\rho)=[\rho]_\sim \in
\mathcal{X}_*$). Our Physical Subsystem can thus be identified with $\mathfrak{X}=(\mathcal{B}(\mathcal{H}_A),\mathcal{B}(\mathcal{H}),P_0)$.

To discuss the dynamics, suppose $\sigma_\beta$ is the bath
density matrix at thermal equilibrium at inverse temperature
$\beta$, and suppose we are given a hamiltonian on $\mathcal{H}$
of the form
\begin{equation}
H_\lambda=H_A\otimes 1+ 1\otimes H_B+ \lambda H_I
\end{equation}
with interaction
\begin{equation}
\lambda H_I= Q\otimes \Phi
\end{equation}
where $Q$ and $\Phi$ are (bounded) self-adjoint operators on the
respective spaces. Since $[H_B,\sigma_\beta]=0$ it follows
that $[U_t,P_0]=0$. Name the bath correlation function as
\begin{equation}
h(t)=\mathrm{Tr}(\sigma_\beta e^{i H_B t}\Phi e^{-i H_B t}\Phi)
\end{equation}
and let
\begin{equation}
\hat{h}(\omega)=\int {dt\over \sqrt{2\pi}} e^{i\omega t}
h(t)
\end{equation}
denote its Fourier transform. Call also
\begin{equation}
\overline{h}= \mathrm{Tr}(\sigma_\beta \Phi)=\mathrm{Tr}(\sigma_\beta \Phi_t), \qquad t\in\mathbb{R}
\end{equation}
the first order contribution.
%We assume $\mathrm{Tr}(\sigma_\beta
%H_B)=0$, which is valid in case $H_B$ is the seat of a quasi free
%representation of the Canonical Anticommutation Relations (CARs)
%with cyclic vector $|\Omega\rangle$, where $\sigma_\beta=|\Omega\rangle\langle\Omega|$ (see \cite{davies1}).
Then upon identifying
$\mathcal{X}\sim B(\mathcal{H}_A)$ one has $Z_0 X=i[H_A,X]$,
$A_{00}X=i\overline{h}[Q,X]$, and one computes
\begin{eqnarray}
K_{T(\lambda)}X &=&  \int {d\omega\over\sqrt{2\pi}}
(\hat{h}(\omega)-\sqrt{2\pi}\,\overline{h}^2 \delta(\omega)) \left( i\int {d\omega'\over 2\pi \omega'}
[Q_{\omega+\omega',\lambda}^\dagger Q_{\omega+\omega',\lambda},X] \right.
\nonumber \\
&&\left. - {1\over 2}\left\{ Q_{\omega,\lambda}^\dagger
Q_{\omega,\lambda}, X\right\} + Q_{\omega,\lambda}^\dagger X Q_{\omega,\lambda} \right) .
\end{eqnarray}
Here we have defined the operators $Q_{\omega,\lambda}$ as
\begin{equation}
Q_{\omega,\lambda}=\sqrt{1\over \sqrt{\pi} T(\lambda)} \int dt\;
e^{-t^2/2T(\lambda)^2} e^{i\omega t} Q_t
\end{equation}
and $Q_t=U_{-t}(Q)$ is the interaction picture subsystem
hamiltonian at time $t$. Note that the presence of first order
contributions manifests itself in that, with physical and diagrammatical terminology, only "connected" correlation functions appear: the term $\hat{h}(\omega)-\sqrt{2\pi}\,\overline{h}^2 \delta(\omega)$ is in fact related, through Fourier Transform, to the connected bath correlation function
\begin{equation}
\mathrm{Tr}(\sigma_\beta (\Phi_{t_1}-\overline{h}1)(\Phi_{t_2}-\overline{h}1))= h(t_1-t_2)-\overline{h}^2.
\end{equation}
This generalizes the example in \cite{davies1} in that the "atom"
in system $A$ needs not be $N$-dimensional, nor does $Z_0$ need to
have discrete spectrum, and runs completely counter the common
idea that a "small subsystem" is implemented by its Hilbert space
being finite dimensional. Rather, the subsystem is small if the
boundedness hypothesis (\ref{eq:no_oscillation_lambda_second}) and (\ref{hp:lambda_convergence_second})
hold true, indicating a fast information flow (faster then second order) from the Physical Subsystem to the remaining degrees of freedom (see \cite{davies1} for conditions on the correlation function $h$ to satisfy these hypotheses).

Note also that this example generalizes the preexisting literature in that it takes first
order corrections into account.

As a final important observation, we note that if $\mathcal{H}_A$ is finite dimensional, then if suitable ergodic conditions on the projected perturbation $Q$ are satisfied (see~\cite{steady}) we know that for all real $\lambda$ there exists a unique steady state $\rho^\lambda_\infty$ for the dual of $W^\lambda_t=\exp\{(Z_0+\lambda A_{00}+\lambda^2 K_{T(\lambda)})t\}$ in $\mathcal{X}_*\sim\mathcal{T}(\mathcal{H}_A)$, and that this state is faithful. If moreover we assume $A_{00}=0$, the steady state analysis in \cite{davies1}, plus a straightforward use of Laplace transforms, imply that
\begin{equation}
\lim_{\lambda\rightarrow 0} \rho^\lambda_\infty={e^{-\beta \mathcal{H}_A} \over \rm{Tr} \left( e^{-\beta \mathcal{H}_A} \right)}
\end{equation}
gives the density matrix at thermal equilibrium, at the heat bath inverse temperature $\beta$. This obviously leads one to conjecture that although the limit dynamics may not be defined at $\lambda=0$, still the limit steady state(s) $\lim_{\lambda\rightarrow 0} \rho^\lambda_\infty$ may exist and provide important physical information (uniqueness being linked to the question of phase transition).

\section*{Summary and Conclusion}
In a recent paper by us we have found a contraction semigroup able to correctly approximate a projected and perturbed one-parameter group of isometries in a generic Banach space, in the limit of weak-coupling. Here we have specialized to what we have defined to be a Physical Subsystem, where the projected subspace is a $W^*$-subalgebra, given in terms of a (completely positive) projecting normal conditional expectation. With this very general and natural requirement we have proven that the contraction semigroup specializes to a Quantum Dynamical Semigroup, thus guaranteeing all important physical properties like positivity and trace conservation.

After reporting a fairly general class of Physical Subsystems, according to our definition, we have discussed the limit dynamics of two important and new examples, the first being a Quantum version of the celebrated Fermi Golden Rule, and the second being the partial tracing over a thermal bath, in case of a general spectrum for the free subsystem dynamics, as well as in the presence of first order terms.

Our results are of very general nature, and open the way to the study of a variety of different extensions (beyond second order, time dependent free dynamics, Boltzmann equation, to name a few) and applications in the fields of continuous variables quantum open systems, quantum information, phase transitions and mesoscopic physics, and allow future investigations of irreversibility and decoherence in more abstract quantum theories.

\section*{Acknowledgements}
We wish to thank Prof. Fausto Rossi (NTL, Phys. Dept., Politecnic of Turin) for profound and enlightening discussions on all the important concepts in this work. We would like to thank Prof. Hisao Fujita Yashima (Dept. Mathematics, University of Turin) for having offered so many days of invaluable help and discussions to the author.

\end{document}